\documentclass{iaus}
\usepackage{graphicx}

\pubyear{2011} 
\volume{284}  
\pagerange{1--12}
\jname{The Spectral Energy Distribution of Galaxies}
\editors{R.J. Tuffs \&  C.C.Popescu, eds.}
\begin{document}

\title{SED fitting with MCMC: 
methodology and application to large galaxy surveys}

\author{Viviana Acquaviva$^1$, Eric Gawiser$^1$ and Lucia Guaita$^2$}

\affiliation{$^1$ Department of Physics and Astronomy, Rutgers, The State University of New Jersey, Piscataway, NJ 08854 
\\[\affilskip] $^2$ Institutionen f{\"o}r Astronomi, Stockholms Universitet, SE-106 91 Stockholm, Sweden \\[0.2cm] email: {\tt vacquaviva@physics.rutgers.edu}}
\maketitle

\begin{abstract}
We present GalMC (Acquaviva et al 2011), our publicly available Markov Chain Monte Carlo algorithm for SED fitting, show the results obtained for a stacked sample of Lyman Alpha Emitting galaxies at z $\sim$ 3, and discuss the dependence of the inferred SED parameters on the assumptions made in modeling the stellar populations. We also introduce SpeedyMC, a version of GalMC based on interpolation of pre-computed template libraries. While the flexibility and number of SED fitting parameters is reduced with respect to GalMC, the average running time decreases by a factor of 20,000, enabling SED fitting of each galaxy in about one second on a 2.2GHz MacBook Pro laptop, and making SpeedyMC the ideal instrument to analyze data from large photometric galaxy surveys.

\end{abstract}

\firstsection 

\section{SED fitting with MCMC: a two-step process}
SED fitting is the process of extracting information on the physical properties of galaxies, such as stellar population age, mass, star formation rate, dust content, metallicity, and redshift, starting from a set of templates that predict how galaxy spectra look like as a function of these properties, which are the SED fitting parameters. This process relies on the simple but powerful idea that since the properties of the models are known, if we can find models that resemble the observations we can infer the properties of the data.
A more rigorous way of comparing models with observations -- in other words, deciding whether a model resembles the data or not - is the $\chi^2$ statistics. For each set of parameters, we need to compute the prediction of what the observations would be if that model was the true one. This step is conceptually simple but complicated in practice, because of the many astrophysical processes that needs to be modeled. GalMC implements this process through the sequence described in Fig. 1.
GalMC is based on Bayesian statistics. Therefore, a second step of the inference process requires to reconstruct the probability distribution of the SED fitting parameters, which are treated as random variables. This is done by exploring the parameter space with a random walk biased so that the frequency of visited locations is proportional to the probability density function. This path through parameter space is the Markov Chain. Once these probabilities are known, one can compute the desired credible intervals for each of the parameters; and because of how visited locations are chosen, integrating the probability distribution functions (PDFs) becomes a simple matter of summing over the points in the chains.

\section{Probability distributions, degeneracies, and the impact of systematics}
Detailed results for two stacked samples of Lyman Alpha Emitting galaxies at $z \sim 3$ were presented in \cite[Acquaviva et al (2011)]{2011arXiv1101.2215A}. Here we just show that GalMC is able to capture multi-modal probability distributions, such as the double peak in the Age vs Stellar Mass distribution, which is due to the degeneracy between these two parameters. We also want to highlight the impact of the assumptions made in modeling the stellar populations, shown for the case of Stellar Mass in the right panel of Fig. 2. The different curves all refer to models commonly used in the literature: the BC03 (\cite{BC03}) and CB07 (\cite{CB07}) stellar population templates, at Solar or variable metallicity, and with or without including nebular emission. The corresponding scatter in the estimate of stellar mass (which does not include the possibility of different initial mass functions, IMFs) is a factor of $\sim$ 2.5, significantly larger than the statistical uncertainty for the same data.
\begin{figure}[t]
\begin{center}
\includegraphics[width=\linewidth]{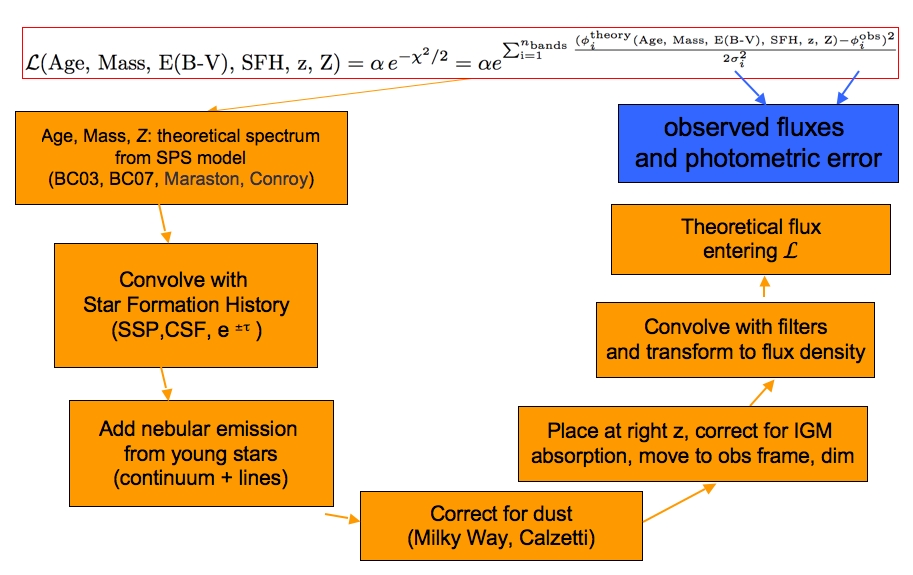}
\caption{The series of steps performed by GalMC to obtain the predicted spectrum as a function of the SED parameters. After the convolution with filters transmission curves, this quantity can be directly compared to the data to obtain a $\chi^2$ value.  }
\end{center}
\end{figure}

\section{SpeedyMC: MCMC for large galaxy catalogs}
MCMC algorithms are much more efficient ways of exploring high-dimensional parameter spaces with respect to algorithms where the probability distribution is sampled at a set of fixed locations on a grid. In fact, the ``interesting" region of parameter space (the one where data and models look like each other) often occupies a small fraction of the total volume. While grid-based models need to explore all of it, Markov Chains are able to ``recognize" the interesting regions and will spend most of the time visiting (sampling) those locations. Yet, the complicated process described in Fig. 1, which leads to the computation of the $\chi^2$ value corresponding to a set of parameters, usually needs to be repeated tens of thousands of times. The computational bottlenecks in this case are the generation of a stellar population template at the right age, and the convolution with the filter transmission curves. To alleviate the first problem, GalMC uses our modified version of GALAXEV (Bruzual and Charlot 2011), developed in collaboration with the authors, which is $\sim$ 20 times faster than the official release. However, the typical time per iteration is still about 0.4 seconds on a 2.2GHz MacBook Pro laptop (for simplicity, all quoted running times will be referred to this machine), and therefore the typical chain per object takes a few hours to run. This becomes impractical for catalogs comprising thousands of objects. The basic idea of SpeedyMC is to find a different (faster) way to compute the $\chi^2$ corresponding to a certain set of parameters. To achieve this objective, we take the following four steps:
\begin{figure}[t!]
\begin{center}
\includegraphics[width=5cm]{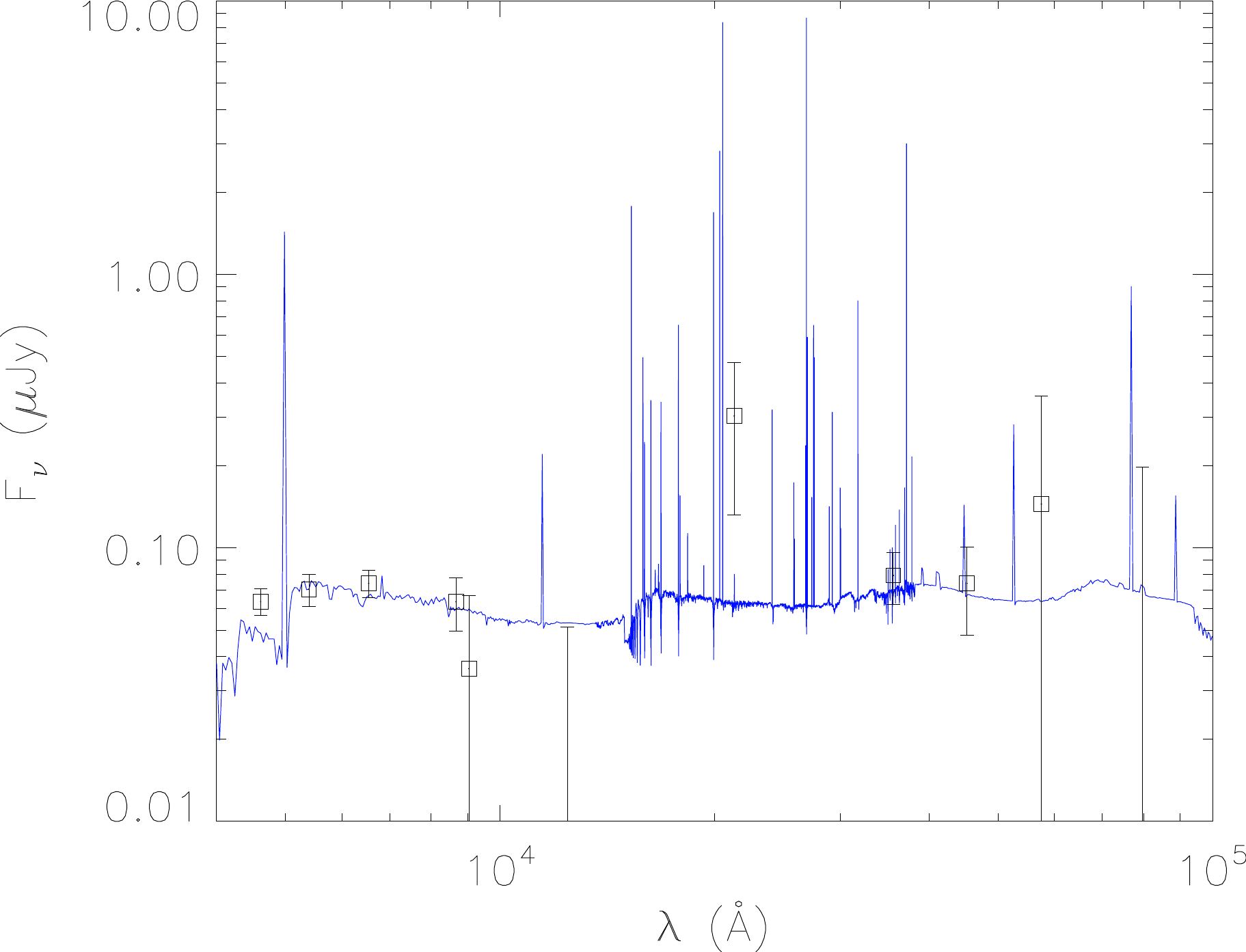}
\includegraphics[width=4.2cm]{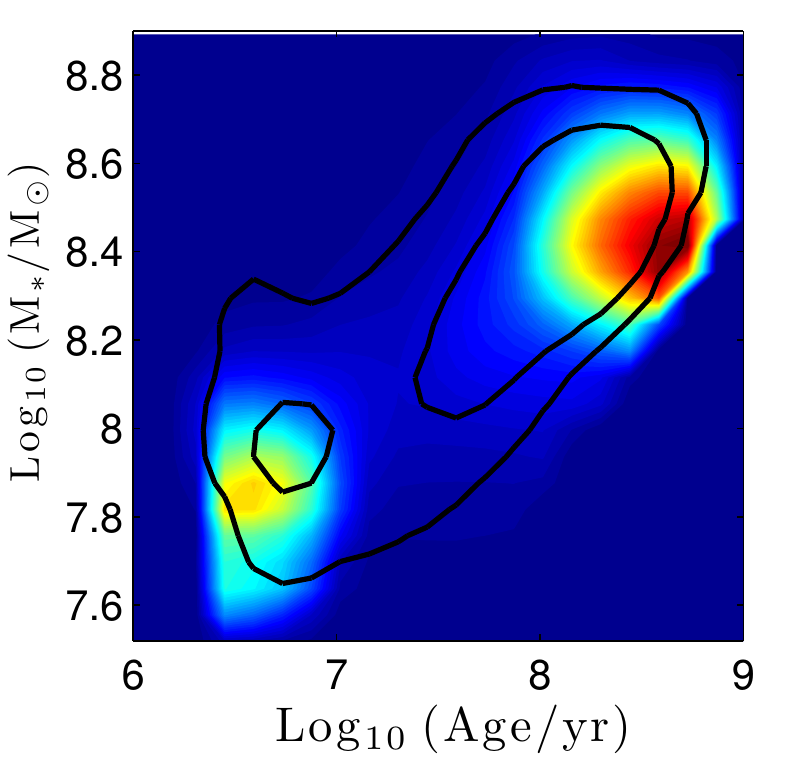}
\includegraphics[width=3.56cm]{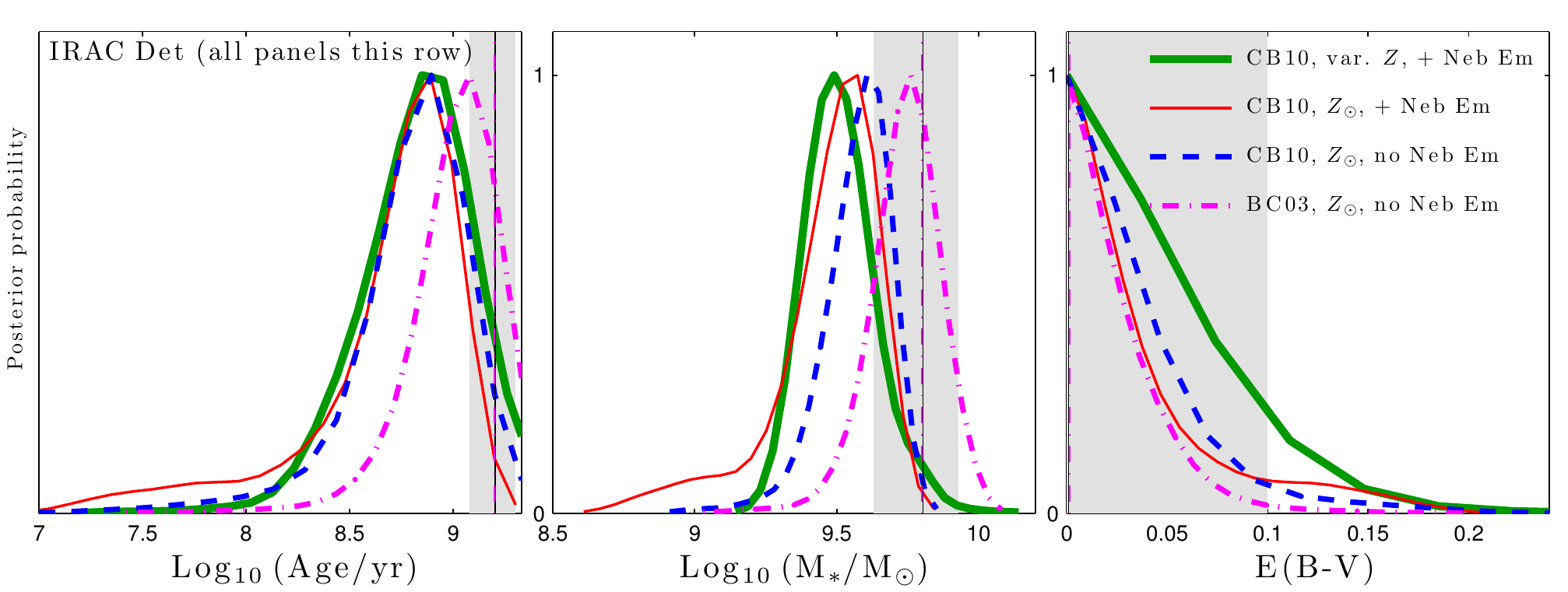}
\label{sed}
\caption{\small{ {\it Left}: Data and best-fit model of the z = 3.1 Lyman Alpha Emitters from \cite{2011arXiv1101.2215A}. {\it Middle}: Marginalized constraints on age and stellar mass. The contours indicate the 68\% and 95\% credible regions, while the color gradient is based on average likelihood in the binned chain. For flat priors, 
lack of exact overlap indicates that the posterior distribution is non-Gaussian; in this case, the contours also show a bi-modal probability distribution. Markov chains are analyzed using the public software from \cite[Lewis and Bridle 2002]{lb}. {\it Right}: Probability distribution for the Stellar Mass assuming $Z = Z_\odot$ for the BC03 (dotted-dashed, magenta) and BC07 (dashed, blue) models, then including nebular emission (thin solid, red), and varying $Z$ with a logarithmic prior (thick solid, green). Shaded regions show the constraints from \cite[Lai et al (2008)]{lai}. }}
\end{center}
\end{figure}

\begin{enumerate}
\item We compute the spectra on a grid of locations exploring the entire parameter space, saving the final product of the sequence of steps described in Fig 1 ({\it after} convolution with the filter transmission curves, so we retain only a handful of numbers corresponding to the flux densities in the observations' bands);
\item We read the grid into memory;
 \item We run MCMC as usual, but to compute the $\chi^2$ at each location we use {\it multi-linear interpolation} between the pre-computed spectra;
 \item We enjoy the speed up factor of 20,000, which allows us to fit the SED of each galaxy in a few seconds (even assuming to run several chains per object).
\end{enumerate}
A couple of caveats are in order. First, this method doesn't have the flexibility of GalMC; because it is difficult to perform interpolation in more than three dimensions, not more than four SED fitting parameters can be used (the fourth parameter being stellar mass, which is a normalization and therefore is excluded from the interpolation process). Second, there is an ``overhead" cost in computing the grid; for 50 values of age and E(B-V), and 100 values of redshift, running the initial grid takes about 24 hours, and this need to be repeated for a different survey (since the set of utilized filters change), or to use, \eg, a different star formation history or IMF. Still, for large surveys the set of filters is fixed, the number of different modeling options one might want to try is limited, and four parameters are enough to capture the general physical properties of a population of galaxies. Finally, let us observe that the resolution of the initial grid {\it does not} correspond to the resolution with which the PDF is sampled, as is the case in grid-based models. MCMC is still free to sample any desired location in parameter space, and the accuracy of the predicted spectrum corresponds to the accuracy of the linear interpolation between the points of the grid. This is illustrated in Fig. 3. The accuracy can be improved by increasing the number of points in the grid. A test conducted on the LAEs at $z = 3.1$ revealed that 50 points in age between 0 and the age of the Universe and 50 values of E(B-V) between 0 and 1 are enough to produce the estimate and credible intervals as the original GalMC, and using 100 values rather than 50 does not produce any appreciable difference.

SpeedyMC is currently being used for the analysis of data from the Cosmic Assembly Near-Infrared Deep Extragalactic Legacy Survey (CANDELS),  (\cite[Grogin et al 2011]{grogin}, \cite[Koekemoer et al 2011]{koekemoer}, \cite[Acquaviva et al 2012]{Acq2012}). The algorithm is not yet public, but you are welcome to contact the author for discussion on how to implement it, starting from GalMC.

\begin{figure}[t]
\begin{center}
\includegraphics[width=8cm]{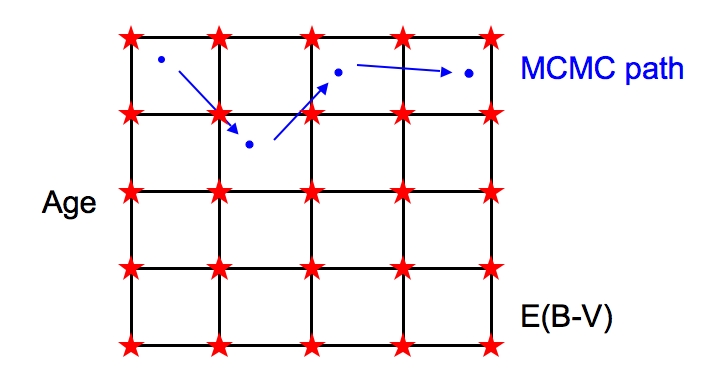}
\caption{An example of the path of SpeedyMC for a two-dimensional grid. The visited locations do not need to lie at the locations where template spectra have been saved (indicated by stars); instead, the corresponding spectrum is obtained by very fast bi-linear interpolation between the four corner stars.  }
\end{center}
\end{figure}

\end{document}